\documentclass[prd,preprint,eqsecnum,nofootinbib,amsmath,amssymb,
               tightenlines,dvips]{revtex4}
\usepackage{graphicx}
\usepackage{bm}
\input epsf

\begin {document}


\def\ca{C_{\rm A}}

\def\cf{C_{\rm F}}

\def\df{d_{\rm F}}
\def\da{d_{\rm A}}
\def\Nf{N_{\rm f}}
\def\Nc{N_{\rm c}}
\def\ta{t_{\rm A}}
\def\tf{t_{\rm F}}
\def\md{m_{\rm D}}
\def\half{\tfrac{1}{2}}
\def\Real{\operatorname{Re}}
\def\alphas{\alpha_{\rm s}}
\def\p{{\bm p}}
\def\q{{\bm q}}
\def\k{{\bm k}}
\def\x{{\bm x}}
\def\B{{\bm B}}
\def\b{{\bm b}}
\def\f{{\bm f}}
\def\F{{\bm F}}
\def\h{{\bm h}}
\def\r{{\bm r}}

\def\qhat{\hat{\bar q}}
\def\grad{{\bm\nabla}}
\def\tr{\operatorname{tr}}
\def\sech{\operatorname{sech}}



\title
    {
      Simple Formula for High-Energy Gluon Bremsstrahlung in a Finite,
      Expanding Medium
    }

\author{Peter Arnold}
\affiliation
    {%
    Department of Physics,
    University of Virginia, Box 400714,
    Charlottesville, Virginia 22904, USA
    }%

\date {\today}

\begin {abstract}%
{%
Previous authors have considered the problem of the medium effects on
single gluon bremsstrahlung associated with producing a high-energy
particle in a finite, time-dependent QCD plasma.
Working to leading logarithmic order,
I show that the result for the bremsstrahlung gluon spectrum
can be cast into a remarkably simple form
in the general case.  I similarly analyze the process of pair
production.  Also, I comment on the radius of convergence
of the opacity expansion in cases where the leading-log approximation
holds, showing that the opacity expansion does not converge when
the thickness of the plasma is greater than roughly the
bremsstrahlung formation time.
Additionally,
as a special bonus---available for a limited time only while supplies
last!---I
summarize translations between the notation used
by a few of the groups who have worked on this and related problems.
}%
\end {abstract}

\maketitle
\thispagestyle {empty}


\section {Introduction and Main Result}
\label{sec:intro}

Roughly a decade ago, Baier, Dokshitzer, Mueller, and Schiff (BDMS)
\cite{BDMS} derived a
simple result for the effect of the medium on the the probability of
single gluon bremsstrahlung from a high-energy parton produced by some
hard process in the background of a uniform, time-independent chunk of hot QCD
matter (known as a ``brick'').
Their simple result (based on application of a more general
formalism) was derived for cases where the number $N_{\rm coh}$ of
coherent soft scatterings during gluon bremsstrahlung is large, and they
looked for a
result valid to leading order in $(\ln N_{\rm coh})^{-1}$.
They found
\begin {equation}
   \omega \, \frac{d}{d\omega}(I-I_{\rm vac}) =
   \frac{\alpha}{\pi} \, x \, P_{s{\to}{\rm g}}(x) \,
   \ln\left| \cos(\omega_0 L) \right| ,
\label {eq:brick}
\end {equation}
where $I$ is the probability of gluon bremsstrahlung from the high-energy
particle of energy $E$ and species $s$ (quark or gluon),
$I_{\rm vac}$ is the corresponding probability had the hard particle
been produced in vacuum,
$\omega = xE$ is the energy of the bremsstrahlung gluon,
$P_{s{\to}{\rm g}}(x)$ is the usual vacuum
splitting function,
$L$ is the distance the high-energy particle travels through the
(uniform) medium before abruptly exiting into vacuum,
and $\omega_0$ is a complex number with phase $\exp(-i\pi/4)$ given
by
\begin {equation}
   \omega_0^2 =
   - i \,
   \frac{[(1-x)\ca + x^2 C_s] \, \qhat}{2 x (1-x) E} \,.
\label {eq:omega}
\end {equation}
Here, $C_R \qhat$ is the average squared transverse momentum transfer
per unit length that a high-energy particle with color representation
$R$ picks
up through soft, elastic collisions with the medium, evaluated at leading-log
order,
\begin {equation}
   \qhat \equiv
   \int d^2q_\perp \>
   \frac{d\bar\Gamma_{\rm el}}{d^2q_\perp} \,
   q_\perp^2 ,
\label {eq:qhat}
\end {equation}
where $C_R \bar\Gamma_{\rm el}$ is the collision rate (which is the same at
leading order for high-energy quarks and gluons, except for an overall
factor of the quadratic color Casimir $C_R$).
The leading-log approximation arises from the need
to cut off the large $q_\perp$ behavior of this integral, which I will
briefly review later.

In another paper \cite{BDMSc},%
\footnote{
   \label{foot:problem}
   Readers should beware that Ref.\ \cite{BDMSc} investigates
   a slightly different problem than the one proposed here, and gets
   a correspondingly different answer, for example, for the brick
   case (\ref{eq:brick}).
   Here, as in Ref.\ \cite{BDMS}, I consider radiation from
   a high-energy parton
   after it leaves a hard collision that occurs inside the medium.
   Ref.\ \cite{BDMSc}, in contrast, purports to study the case where
   the particle approaches the medium from the outside.
   See the discussion immediately following Eq.\ (42b) of
   Ref.\ \cite{BDMS}.
}
BDMS showed that they could also find
leading-log results for non-uniform, time-dependent media, such as
an expanding quark-gluon plasma.  The result was not as simple,
however, involving a double integral of a complicated function
found for the particular case they studied.  In this paper, I show
that there is a magically simple generalization of
(\ref{eq:brick}) to the general case of non-uniform, time-dependent
media.  The result is
\begin {equation}
   \omega \frac{d}{d\omega}(I-I_{\rm vac}) =
   \frac{\alpha}{\pi} \, x \, P_{s{\to}{\rm g}}(x) \,
   \ln\left| c(0) \right| ,
\label {eq:wow}
\end {equation}
where $c(t)$ satisfies the differential equation
\begin {equation}
   \frac{d^2c}{dt^2} = - \omega_0^2(t) \, c(t)
\label {eq:c}
\end {equation}
with the boundary condition that $c(t)$ approach the constant 1
as $t \to \infty$, and the convention that $t=0$ is the time of
the hard collision that produced the initial high energy particle.
Here, $\omega_0^2(t)$ is (\ref{eq:omega}) evaluated at the position
of the high-energy particle at time $t$,
and now $\qhat = \qhat(t,\x(t))$ is time dependent.
The fact that the particle eventually ends up in vacuum means
that $\omega_0^2(t) \to 0$ as $t\to\infty$.

I will later give the generalization of the result to the
case ${\rm g}\to{\rm q}\bar{\rm q}$ of pair production.

I should note that
BDMS's result and my generalization are not complete descriptions
of the average bremsstrahlung spectrum at leading-log order
\cite{ZakharovResolution}.  For sufficiently small $L$,
the average medium effect on bremsstrahlung is instead dominated by
atypical events where there is a single, larger-than-normal
scattering from the medium.  I will review this later, along
with the condition on $L$ \cite{N1}.

The simple form (\ref{eq:wow}) is peculiar to
three spatial dimensions (i.e.\ two transverse dimensions). I do not
know of a generalization that would give a comparably simple result in
other dimensions.

In the next section, I review the starting point for the calculation,
based on the formalism of Zakharov \cite{Zakharov2} and
Baier, Dokshitzer, Mueller, Peigne, and Schiff (BDMPS)
\cite{BDMS,BDMPS1,BDMPS2,BDMPS3}.  I organize the notation in a way
that's a little friendlier
for perturbative calculations in a QCD medium with non-static
scatterers than the original BDMPS version.
(See the discussion in the appendix.)
Then I review the leading log approximation and the range
of validity of the BDMS result (\ref{eq:brick}).
In section \ref{sec:derivation}, I derive the basic result
(\ref{eq:wow}) of this paper.
Section \ref{sec:examples} then gives various examples for some
cases where the equation (\ref{eq:c}) for $c(t)$ has analytic
solutions.  Section \ref{sec:limits} analyzes the general
problem in the limiting cases of a QCD medium that is narrow or
wide compared to the formation length for gluon bremsstrahlung.
Throughout this paper, I focus on the case of bremsstrahlung in
order to simplify notation, but the formalism applies equally well
to pair production.  In section \ref{sec:pair}, I give the
corresponding results for the case of pair production.
Finally, in section \ref{sec:opacity}, I comment on implications of
BDMS's original result (\ref{eq:brick}) for the convergence of what
is know as the opacity expansion---the expansion of the bremsstrahlung
probability in powers of the number of elastic scatterings.

The notational conventions that I use are not exactly the same as
those of BDMS or Zakharov.  The relationship between my notation and various
other authors is discussed in Appendix \ref{app:notation}.




\section {Starting Point and Assumptions}

\subsection {Notational preliminaries}

Throughout, I will use $C_R$ to denote the quadratic Casimir
$T_{R}^a T_{R}^a$ for the color representation $R$ associated
with some particle, with color generators $T_R^a$.
For a particle of type $s$, I will abbreviate
this as $C_s$.  For QCD,
\begin {equation}
   C_{\rm q} \equiv \cf = \frac{N_c^2-1}{2N_c} = \frac43 ,
   \qquad
   C_{\rm g} \equiv \ca = N_c = 3 ,
\end {equation}
where $N_c=3$ is the number of colors.
$d_R$ will be the dimension of the color representation, so that
\begin {equation}
   d_{\rm q} \equiv \df = N_c = 3,
   \qquad
   d_{\rm g} \equiv \da = N_c^2-1 = 8.
\end {equation}
$t_R = C_R d_R/d_A$ will be the trace normalization defined by
$\tr(T_R^a T_R^b) = t_R \delta^{ab}$, with
\begin {equation}
   t_{\rm q} \equiv t_{\rm F} = \half \,,
   \qquad
   t_{\rm g} \equiv t_{\rm A} = N_c = 3 .
\label {eq:ts}
\end {equation}
The
Dokshitzer-Gribov-Lipatov-Altarelli-Parisi (DGLAP)
splitting functions in (\ref{eq:brick}) and (\ref{eq:wow}) are
\begin {align}
   P_{{\rm q}\to {\rm g}}(x)
   &= \cf \, \frac{[1+(1-x)^2]}{x} \,,
\\
   P_{{\rm g}\to {\rm g}}(x)
   &= \ca \, \frac{[1 + x^4 + (1-x)^4]}{x(1-x)} \,.
\end {align}

Throughout this paper, I will generally place a bar over quantities
when I have removed an overall
factor of $C_R$ from its definition.
So I work with $\qhat$, for example, instead of the more standard
$\hat q$.


\subsection {General Formalism}
\label {sec:general}

Calculations of bremsstrahlung from sufficiently high energy jets must
take into account the Landau-Pomeranchuk-Migdal (LPM) effect, which
arises when the quantum mechanical duration (formation time) of the
bremsstrahlung process becomes comparable to, or exceeds, the
mean free time for small-angle elastic collisions.
The basic procedure for making such calculations was laid out for QED
by Migdal in 1956 \cite{Migdal}.
The generalization to QCD requires accounting for
the fact that a bremsstrahlung gluon, unlike a photon, carries (color)
charge and so can also undergo collisions during the formation time.
I will find it convenient to start with the particular version of this result
derived by Zakharov \cite{Zakharov1,Zakharov2}.
This is equivalent to the BDMPS
formalism of Baier {\it et al.} \cite{BDMS,BSZ}, and I will use
some of that correspondence in how I choose to write Zakharov's
result.
The general formula is
\begin {multline}
   \omega \, \frac{d}{d\omega}(I-I_{\rm vac}) =
   \frac{\alpha x \, P_{s{\to}{\rm g}}(x)}{[x(1-x) E]^2} \,
   \Real
   \int_0^\infty dt_1 \int_{t_1}^\infty dt_2 \>
\\
   \Bigl[
     \grad_{\B_1} \cdot \grad_{\B_2}
     \bigl\{
       G(\B_2,t_2;\B_1,t_1)
       -
       G_{\rm vac}(\B_2,t_2;\B_1,t_1)
     \bigr\}
   \Bigr]_{B_1=B_2=0} ,
\label {eq:general}
\end {multline}
where $G(\B_2,t_2;\B_1,t_1)$ is the Green's function for a
two-dimensional quantum mechanics problem with the
time-dependent Hamiltonian
\begin {equation}
   H(t) = \delta E(\p_B,t) - i \Gamma_3(\B,t) .
\label {eq:H}
\end {equation}
The two terms in $H$ above will be described in a moment.
The Green's function $G(\B,t;\B_1,t_1)$ is a solution to the
Schr\"odinger equation
\begin {equation}
   i \partial_t \psi(\B,t) = H(t) \, \psi(\B,t)
\label {eq:schro}
\end {equation}
with initial condition
\begin {equation}
   G(\B,t_1;\B_1,t_1) = \delta^{(2)}(\B-\B_1) .
\end {equation}

The first term in (\ref{eq:H}) describes the energy difference
\begin {equation}
   (E_{s,\p} + E_{{\rm g},\k}) - E_{s,\p+\k}
   \simeq
   \frac{p_\perp^2 + m_s^2}{2p}
   + \frac{k_\perp^2 + m_{\rm g}^2}{2k}
   - \frac{|\p_\perp+\k_\perp|^2 + m_s^2}{2(p+k)}
\end {equation}
between (i) a high-energy parton of momentum ${\bm P} = \p+\k$ and
energy $E = P$ and (ii)
the same parton with momentum $\p$ plus a bremsstrahlung gluon with
momentum $\k$.  If (following Ref.\ \cite{BDMS}) one defines
\begin {equation}
   \p_B \equiv \frac{p \k_\perp - k \p_\perp}{P} \,,
\end {equation}
then this energy difference
can be rewritten as
\begin {align}
  \delta E(\p_B,t)
  &\equiv
    \frac{p_B^2}{2 x (1-x) P}
    + \frac{m_s^2(t)}{2(1-x)P}
    + \frac{m_{\rm g}^2(t)}{2xP}
    - \frac{m_s^2(t)}{2P} 
\nonumber\\
  &=
    \frac{p_B^2 + x^2 \, m_s^2(t) + (1-x) \, m_{\rm g}^2(t)}{2 x (1-x) E}
  \,.
\label {eq:deltaE}
\end {align}
The notation $m(t)$ accounts for the fact that the effective masses
will change
as the particle transverses a inhomogeneous or time-dependent medium.
Qualitatively, the expectation of $1/\delta E(\p_B)$ is
of order the formation time for the bremsstrahlung process
in the medium.%
\footnote{
   There is a difference between my use of
   the phrase ``formation time'' and Zakharov's
   \cite{Zakharov2}.
   See Appendix \ref{app:notation}.
}
The second term in (\ref{eq:H}) is given by
\begin {equation}
  \Gamma_3(\B,t) =
  \half \ca \, \bar\Gamma_2(\B,t)
  + (C_s - \half\ca) \, \bar\Gamma_2(x\B,t)
  + \half\ca \, \bar\Gamma_2\bigl((1-x)\B,t\bigr) ,
\label {eq:Gamma3}
\end {equation}
where $\bar\Gamma_2$ is related to the Fourier transform of
$d\bar\Gamma_{\rm el}/d^2q_\perp$ and defined by
\begin {equation}
   \bar\Gamma_2(\b,t) \equiv
   \int d^2 q_\perp \>
   \frac{d\bar\Gamma_{\rm el}(t)}{d^2q_\perp}
   \,
   (1 - e^{i \b \cdot \q_\perp})
   =
   \frac{1}{\pi}
   \int d^2 q_\perp \>
   \frac{d\bar\Gamma_{\rm el}(t)}{d(q_\perp^2)}
   \,
   (1 - e^{i \b \cdot \q_\perp}) .
\label {eq:Gamma2}
\end {equation}

I have not used exactly the same notation as either Zakharov
or BDMS, and I summarize the differences of notation in Appendix
\ref{app:notation}.  On a slightly more substantive matter,
both implicitly assumed that
the rate $\Gamma_{\rm el}$ for soft scattering of the high-energy
particle could be written as a number density $n$ of static particles in
the medium times a cross-section $\sigma_{\rm el}$ for scattering
from such particles.  However, their results do not actually depend
on this assumption.  If one simply writes their formulas in terms of
the rate $\Gamma_{\rm el}$ rather than $n \sigma$,
then they apply equally well to the
case of scattering from non-static particles, which, for example,
was analyzed for leading-order calculations in an infinite,
time-independent thermal medium by Arnold, Moore, and Yaffe (AMY)
\cite{AMYsansra,AMYkinetic,AMYx}
and Jeon and Moore \cite{JeonMoore}.  Specifically,
the differential rate is
\begin {equation}
   \frac{d\bar\Gamma_{{\rm el},s}}{d^2q_\perp}
   =
   \int dq_z
   \sum_{s_2} \nu_{s_2}
   \int \frac{d^3 p_2}{(2\pi)^3} \>
   \frac{d\bar\sigma_{\rm el}}{d^3q}
   \,
   f_{s_2}(\p_2) \, \bigl[ 1 \pm f_{s_2}(\p_2-\q) \bigr] ,
\end {equation}
Here, $C_R \bar\sigma_{\rm el}$ is the soft, elastic scattering rate for
a high-energy particle to scatter from a single plasma particle of
momentum $\p_2$ and species $s_2$.
$\q_\perp$ is the transverse momentum transfer to the high-energy particle from
this single scattering.
$f(\p_2)$ is the phase space
density of plasma particles per degree of freedom, which in
thermal equilibrium is the Bose or Fermi distribution for the
plasma particle.
$\nu_{s_2}$ is the number of spin, color, and flavor
degrees of freedom for species $s_2$,
which would be $2 \da = 16$ for gluons and $4 \df \Nf = 12 \Nf$ for
the sum of quarks and anti-quarks, where $\Nf$ is the number of
quark flavors.
The factor of $f$ gives the density of plasma particles, while
the factor of $1 \pm f$ is a final-state Bose or Fermi factor.
Final state factors for the high-energy particle (as opposed
to the plasma particle it is scattering from) may be included at
the end of the LPM calculation, if desired (see, for example,
the $1{\to}2$ splitting terms in the effective kinetic theories of
Refs.\ \cite{AMYsansra,AMYkinetic,bottom_up}), but in the present context
I assume that the high-energy particle is an isolated particle of
energy much higher than the plasma particles, so that its final
state factor can be ignored.

In terms of specifics, perturbative calculations for a QCD plasma
in local equilibrium give the simple
formulas%
\footnote{
  The simple form of the $q_\perp \ll T$ formula comes from
  Ref.\ \cite{AGK}.
  This is the formula used by AMY \cite{AMYx}
  in studying
  the LPM effect in hydrodynamic transport coefficients, where the relevant
  particle energies are $E \sim T$.
}
\begin {equation}
   \frac{d\bar\Gamma_{\rm el}}{d^2q_\perp}
   \simeq
   \frac{1}{(2\pi)^2} \times
   \begin {cases}
      \displaystyle\frac{g^2 T \md^2}{q_\perp^2 (q_\perp^2+\md^2)} \,,
      & q_\perp \ll T, \\[15pt]
      \displaystyle\frac{g^4 {\cal N}}{q_\perp^4} \,,
      & q_\perp \gg T,
   \end {cases}
\label {eq:dsig}
\end {equation}
in the limits of $q_\perp$ small or large compared to the
temperature $T$.
Here $\md$ is the Debye mass,
\begin {equation}
   \md^2 = \Bigl( \ta + \Nf \tf \Bigr)
           \tfrac{1}{3} g^2 T^2
   = \left( 1 + \tfrac16 \Nf \right) g^2 T^2 ,
\label {eq:md}
\end {equation}
and ${\cal N}$ is the weighted number density
\begin {equation}
  {\cal N}
  \equiv
  \sum_{s_2} \nu_{s_2} t_{s_2}
  \int \frac{d^3\p_2}{(2\pi)^3} \>  f_{s_2}(\p_2)
   = \frac{\zeta(3)}{\zeta(2)}
              \Bigl( \ta + \tfrac32 \Nf \tf \Bigr)
              \tfrac13 T^3
   = \frac{\zeta(3)}{\zeta(2)} \left( 1 + \tfrac14 \Nf \right) T^3 ,
\label {eq:N}
\end {equation}
where $\zeta(z)$ is the Riemann zeta function.

The formalism reviewed above assumes that the characteristics of the medium do
not change significantly over a Debye screening length.  It is not
restricted to equilibrium situations,
but I will assume that the differential elastic
cross-section is isotropic in the transverse plane.
The formalism also assumes that the final bremsstrahlung gluon and
accompanying particle are energetic enough that transverse
momentum transfers from the medium will be small compared to their momenta.

More generally, all calculations based on variations of Migdal's
procedure require that the mean free path for soft, elastic collisions
be large compared to the screening length.%
\footnote{
  More precisely, it is the mean free path for the subset of soft, elastic
  collisions which contribute to the result at the desired accuracy.
  In the thermal case, for example, ultra-soft magnetic interactions with
  $q_\perp \sim g^2 T$ do not affect results at
  leading order in coupling.
}
This assumption holds
for a thermal plasma in the weak coupling limit,
where the mean free path is order
$1/g^2T$ and the screening length is order $1/gT$.
(See, for example, the discussion in Ref.\ \cite{AMYsansra}.)

If running of the coupling constant $\alpha$ is included in
the analysis, then
$d\bar\Gamma_{\rm el}/d^2q_\perp$ should plausibly be evaluated with
$g^2(q_\perp)$.%
\footnote{
   See, for example, Refs.\ \cite{Peshier, ArnoldDogan}.
   In order to avoid an unphysical infrared divergence of the
   calculation when the definition of $g^2$ blows up at
   $\Lambda_{\rm QCD}$, one should appropriately
   cut off the running in the infrared.  One possibility would
   be to use $g^2(\sqrt{q_\perp^2 + \md^2})$.
}
In Ref.\ \cite{ArnoldDogan}, it is argued that the overall factor of
$\alpha$ associated with the coupling of the bremsstrahlung gluon
[here the explicit $\alpha$ in (\ref{eq:wow})
or (\ref{eq:general})],
should plausibly be evaluated as
$\alpha(Q_\perp)$, where $Q_\perp$ is the typical transverse momentum
transfer over the formation time and is discussed below.
This last prescription is in the spirit of earlier suggestions
by BDMPS \cite{BDMPS3}.%
\footnote{
   Specifically, after Eq.\ (3.12) of Ref.\ \cite{BDMPS3}, they
   suggest taking $\alphas(k)$ with $k \propto L^{1/2}$
   for the calculation
   of average bremsstrahlung energy loss in a thin QCD medium.
   For that problem, the energy loss is dominated by gluons whose
   formation time is of order the length $L$ of the medium.
   In that case, $Q_\perp \propto L^{1/2}$ as in (\ref{eq:Qperp})
   below.
}


\subsection {Leading Log (Harmonic Oscillator) Approximation}

Consider a medium that is thick enough that the total number of soft
scatterings with individual momentum transfers $q_\perp \gtrsim \md$,
as the particle traverses the medium, is large.%
\footnote{
  This statement contains the restriction $q_\perp \gtrsim \md$
  because, in the weak coupling
  limit, the most common scatterings,
  by a parametric factor of $\ln(\alpha^{-1})$,
  have momentum $g^2 T \lesssim q_\perp \ll \md$
  and are mediated by the exchange of low-frequency magnetic gluons.
  These low-frequency magnetic gluons are not Debye screened,
  and their contribution is cut off only by
  non-perturbative effects.
  However, these ultra-low momentum scatterings do not contribute
  at leading order to (\ref{eq:qhat}) and (\ref{eq:Gamma2}) [because
  of the factor of $q_\perp^2$ in (\ref{eq:qhat})] and
  so do not have an effect on bremsstrahlung at leading order
  in coupling.
}
In the high energy limit, the number $N_{\rm coh}$ of such scatterings
in a formation time also becomes large.
As noted long ago by Migdal \cite{Migdal}, the calculation of the LPM
effect simplifies significantly if one works to leading order in
$(\ln N_{\rm coh})^{-1}$.  In the context of QCD, such leading-log
calculations were explored by Baier {\it et al.}\
using their BDMPS formalism and what is known as the harmonic oscillator
approximation.  Following BDMS \cite{BDMSc},
I will focus on leading-log calculations
in this paper as well.

The large $N_{\rm coh}$ limit corresponds to the case where the
total transverse momentum transfer $Q_\perp$ to a high-energy particle
during the formation time is large compared to the screening
mass $m_{\rm D}$.  One consequence of large transverse momentum
is that we can ignore the effective particle masses $m_s$ and
$m_{\rm g}$ in (\ref{eq:deltaE}).  Another consequence is that
large $Q_\perp$ corresponds in Fourier space to small $B$.
Naively, Eq.\ (\ref{eq:Gamma2}) for $\bar\Gamma_2$ can then be replaced
by its small $b$ limit, which is formally
\begin {equation}
   \bar\Gamma_2(\b,t) \simeq
   \tfrac14
   \int d^2q_\perp \>
   \frac{d\bar\Gamma_{\rm el}(t)}{d^2q_\perp} \,
   q_\perp^2 b^2
   = \tfrac14 \, \qhat(t) \, b^2 .
\label {eq:Gamma2LL}
\end {equation}
This is known as the harmonic oscillator approximation because
of the form of (\ref{eq:Gamma2LL}).
The problem is that the above integral is logarithmically divergent
because of the large $q_\perp$ behavior of (\ref{eq:dsig}).
For a leading log analysis of typical events,
it should be cut off at order of the typical
total momentum transfer $Q_\perp$ in a formation time.
Parametrically, recalling the definition of $\qhat$,
\begin {equation}
   Q_\perp \sim
   \begin {cases}
      (C_{R_i} \qhat L)^{1/2} ,            & L \lesssim L_{\rm cr}, \\
      (C_{R_i} \qhat L_{\rm cr})^{1/2} ,   & L \gtrsim L_{\rm cr};
   \end {cases}
\label {eq:Qperp}
\end {equation}
where $L$ is the characteristic thickness of the medium and
$L_{\rm cr}$ is the infinite-medium formation time%
\footnote{
   See, for example, the discussion in Sec.\ 3 of Ref.\ \cite{BDMPS2}.
}
\begin {equation}
   L_{\rm cr} \sim \left(\frac{E_i}{C_{R_i} \qhat}\right)^{1/2} .
\label {eq:Lcr}
\end {equation}
Above, $E_i$ is the energy $E$, $xE$, or $(1-x)E$ of a particular
parton in the splitting process, and one should use whichever
parton gives the smallest $Q_\perp$.  For small $x$, that will
be the bremsstrahlung gluon, giving
$L_{\rm cr} \sim (\omega/\ca \qhat)^{1/2}$.

Using (\ref{eq:qhat}) and (\ref{eq:dsig}),
the leading-log value of $\qhat$ for a weakly-coupled thermal
QCD plasma is then%
\footnote{
   For (\ref{eq:qhatLLa}), see also Eq.\ (13) of Ref.\ \cite{BMT}
   and the relation to
   Ref.\ \cite{MooreTeaney} discussed after Eq. (61) of Ref.\ \cite{BMT}.
}
\begin {subequations}
\label {eq:qhatLL}
\begin {equation}
   \qhat \simeq \alpha T \md^2 \ln\left(\frac{Q_\perp^2}{\md^2}\right)
\label {eq:qhatLLa}
\end {equation}
if $Q_\perp \lesssim T$ and
\begin {equation}
   \qhat \simeq
   \alpha T \md^2 \ln\left(\frac{T^2}{\md^2}\right)
   + 4\pi \alpha^2 {\cal N} \ln\left(\frac{Q_\perp^2}{T^2}\right)
\end {equation}%
\label {eq:qhatLLb}%
\end {subequations}%
otherwise.
For 3-flavor QCD, $\alpha T \md^2$ and $4\pi\alpha^2{\cal N}$ differ
by only about 15\%, and so one could combine the logarithms of
(\ref{eq:qhatLLb}) into either
$\alpha T \md^2 \ln(Q_\perp^2/\md^2)$ or
$4\pi \alpha^2{\cal N} \ln(Q_\perp^2/\md^2)$ without much error.

If $Q_\perp$ is so large that $\alpha(Q_\perp)$ is significantly
different from $\alpha(\md)$, then one should include
1-loop running of the coupling when integrating (\ref{eq:dsig}).
The result can be put into the form%
\footnote{
  See, for example, the discussion in Sec.\ VI of Ref.\
  \cite{ArnoldDogan}.  Though the form of (\ref{eq:running}) is
  convenient, it can be misleading.  In the limit that $Q_\perp$
  is so large that $\alpha(Q_\perp) \ll \alpha(T)$, the answer
  does not actually depend on $q_\perp$ of order $Q_\perp$ --- it
  is instead dominated by those $q_\perp$ for which
  $\alpha(q_\perp)$ is of order $\alpha(T)$ \cite{ArnoldDogan}.
  Also, the simple formula
  (\ref{eq:running}) is only valid if there are no vacuum mass thresholds
  between $\md$ and $Q_\perp$.
}
\begin {equation}
   \qhat \simeq
   \alpha(T) \, T \, \md^2(\md) \, \ln\left(\frac{T^2}{\md^2}\right)
   + 4\pi \, \alpha(Q_\perp) \, \alpha(T)
     \, {\cal N} \ln\left(\frac{Q_\perp^2}{T^2}\right) ,
\label {eq:running}
\end {equation}
where $\md(\md)$ indicates the Debye mass (\ref{eq:md}) evaluated
with running coupling $g^2(\md)$.

Note that the leading-log formula (\ref{eq:qhatLL}) for $\qhat$
depends logarithmically on $Q_\perp$, which in turn depends on $\qhat$.
One could determine $\qhat$ self-consistently, but it should be
kept in mind that a precise value of $\qhat$ inside the logarithm
is not called for because we are only pursuing a leading-log result.
For an example of how things work out at next-to-leading logarithmic
order, see the infinite-medium calculation of Ref.\ \cite{ArnoldDogan}.

In any case, in the leading-log approximation (\ref{eq:Gamma2LL}),
the 2-dimensional Hamiltonian of (\ref{eq:H}) becomes a 2-dimensional
harmonic oscillator problem with time-dependent frequency:
\begin {equation}
   H(t) \simeq \frac{p_B^2}{2M} + \half M \, \omega_0^2(t) \,  B^2 ,
\label {eq:HLL}
\end {equation}
with
\begin {align}
   M &= x(1-x)E ,
\\
   \omega_0^2(t) &= -i \, \frac{[(1-x) \ca + x^2 C_s] \, \qhat(t)}{2x(1-x)E}
   \,.
\label {eq:omega2}
\end {align}
Note that $\omega_0^2$ is imaginary.
Its inverse magnitude $1/|\omega_0|$ is of order the infinite-medium
formation time $L_{\rm cr}$ of (\ref{eq:Lcr}).

The harmonic oscillator approximation breaks down for sufficiently small
$L$, even when logarithms are large.  Using the typical total momentum
transfer (\ref{eq:Qperp}) as an upper cut-off to determine the integral
in (\ref{eq:Gamma2LL}) ignores the possibility of bremsstrahlung from
rare, atypical scatterings with lager $q_\perp$, which turn out to be
important for sufficiently small $L$.  There has been some
confusion about the resulting range of validity of the harmonic oscillator
approximation used by BDMS for a leading-log analysis of the spectrum.
Zakharov \cite{ZakharovResolution} suggested
that the harmonic oscillator approximation outlined in this section
is only valid when $L \gg L_{\rm cr}$, which is equivalent to
$|\omega_0 L| \gg 1$.
In Ref.\ \cite{N1}, however, I argue that the validity extends to
\begin {equation}
   L \gg \frac{L_{\rm cr}}{[\ln(Q_\perp^2/\md^2)]^{1/2}} \,.
\end {equation}
This includes the interesting region $L \sim L_{\rm cr}$
(equivalently $|\omega_0 L| \sim 1$)
in a leading-log analysis, which treats the
logarithm as large.


\section {Derivation}
\label {sec:derivation}

\subsection {A double integral}

If $G$ is the Green's function, then
the two components of the vector function
$\grad_{\B_1} G(\B,t;\B_1,t_1)$ will also satisfy the
Schr\"odinger equation (\ref{eq:schro}) but with initial condition
\begin {equation}
   \grad_{\B_1} G(\B,t_1;\B_1,t_1) = \grad_{\B_1} \delta^{(2)}(\B-\B_1) .
\end {equation}
In (\ref{eq:general}), we are interested in the particular case
$\B_1 = 0$, which then corresponds to the initial condition
\begin {equation}
   \grad_{\B_1} G(\B,t_1;\B_1,t_1) \Bigl|_{B_1=0}
   = - \grad_{\B} \, \delta^{(2)}(\B) .
\end {equation}
The desired solution in the leading log approximation
(\ref{eq:HLL}) is
\begin {equation}
   \grad_{\B_1} G(\B,t;\B_1,t_1) \Bigl|_{B_1=0}
   = - \frac{M^2}{2\pi \, S^2(t;t_1)} \, \B \,
     \exp\left( \frac{i M \partial_t S(t;t_1)}{2 S(t;t_1)} \, B^2 \right)
   ,
\label {eq:dG}
\end {equation}
where $S(t;t_1)$ satisfies the differential equation
\begin {equation}
   \partial_t^2 S = - \omega_0^2(t) S
\label {eq:Seq}
\end {equation}
with boundary conditions
\begin {equation}
   S(t_1;t_1) = 0,
   \qquad
   \partial_t S(t;t_1) \Bigl|_{t=t_1} = 1.
\end {equation}
One may check this by (i) plugging it into the Schr\"odinger equation
and noting that it is a solution, and (ii) checking the initial condition
by solving for $t$ infinitesimally close to $t_1$, where
$S(t;t_1) \to t-t_1$ and (\ref{eq:dG}) becomes
\begin {align}
   \grad_{\B_1} G(\B,t;\B_1,t_1) \Bigl|_{B_1=0}
   &\to - \frac{M^2}{2\pi (t-t_1)^2} \, \B \,
     \exp\left( \frac{i M}{2 (t-t_1)} \, B^2 \right)
\nonumber\\
   &= - \grad_{\B} \frac{M}{2\pi i (t-t_1)} \,
     \exp\left( - \frac{M}{2i (t-t_1)} \, B^2 \right)
   ,
\end {align}
which is a representation of $- \grad_\B \, \delta^{(2)}(\B)$ for
infinitesimal $t-t_1$.

Substituting (\ref{eq:dG}) into (\ref{eq:general}) gives
\begin {equation}
   \omega \, \frac{d}{d\omega}(I-I_{\rm vac}) =
   \frac{\alpha}{\pi} \, x \, P_{s{\to}{\rm g}}(x) \,
   \Real {\cal I},
\label {eq:result1}
\end {equation}
where
\begin {equation}
   {\cal I} \equiv
   - \int_0^\infty dt_1 \int_{t_1}^\infty dt_2 \>
   \Bigl[
     \frac{1}{S^2(t_2;t_1)} - \frac{1}{(t_2-t_1)^2}
   \Bigr] .
\label {eq:I1}
\end {equation}
This gives an answer in terms of a double integral involving
the function $S(t_2;t_1)$.  But both integrals can be done
explicitly, even for the case of arbitrary $\omega_0^2(t)$.


\subsection {The $t_2$ integration}

Now consider the other solution to the 2nd-order differential
Eq. (\ref{eq:Seq}), which I will call $C(t;t_1)$ and take
to have boundary conditions
\begin {equation}
   C(t_1;t_1) = 1,
   \qquad
   \partial_t C(t;t_1) \Bigl|_{t=t_1} = 0.
\label {eq:Cboundary}
\end {equation}
If $\omega_0^2(t)$ were a constant, then the two solutions would be
$S = \omega_0^{-1} \sin\bigl(\omega_0 (t-t_1)\bigr)$ and
$C = \cos\bigl(\omega_0 (t-t_1)\bigr)$, which is the motivation for the
labels $S$ and $C$.

The form of the differential equation implies
that the Wronskian
\begin {equation}
   W = C \partial_t S - S \partial_t C
\end {equation}
is independent of time and so always equal to its value at $t=t_1$:
\begin {equation}
   C \partial_t S - S \partial_t C = 1 .
\end {equation}
Dividing both sides by $S^2$ then gives
\begin {equation}
   -\partial_t\left(\frac{C}{S}\right) = \frac{1}{S^2} \,.
\end {equation}
We can use this to do the $t_2$ integral in (\ref{eq:I1}).
Rewrite the time integrals in (\ref{eq:I1}) to have upper limit
$t$, taking the limit $t \to\infty$ at the end.
Then rewrite the $t_2$ integral as
\begin {align}
    \lim_{\epsilon\to0}
    \int_{t_1+\epsilon}^t dt_2 \>
    \left( \frac{1}{S^2(t_2;t_1)} - \frac{1}{(t_2-t_1)^2} \right)
    &= - \frac{C(t;t_1)}{S(t;t_1)} + \frac{1}{t-t_1}
       + \lim_{\epsilon\to0} \left[ 
         \frac{C(t_1+\epsilon;t_1)}{S(t_1+\epsilon;t_1)}
         - \frac{1}{\epsilon}
      \right]
\nonumber\\
    &= - \frac{C(t;t_1)}{S(t;t_1)}  + \frac{1}{t-t_1}
    \,.
\end {align}
So
\begin {equation}
  {\cal I}
  = \lim_{t\to\infty}
    \int_0^t dt_1 \> \left[
         \frac{C(t;t_1)}{S(t;t_1)} 
         - \frac{1}{t-t_1}
    \right] .
\label {eq:I2}
\end {equation}


\subsection {The $t_1$ integration}

Now note that
\begin {equation}
   C(t;t_1) = - \partial_{t_1} S(t;t_1) .
\label {eq:Sderiv}
\end {equation}
This follows because (i) $-\partial_{t_1} S$ will satisfy the same
equation (\ref{eq:Seq}) that $S$ does,
and (ii) the boundary conditions work out
correctly.  The boundary conditions (\ref{eq:Cboundary}) can be
confirmed from the small $t-t_1$ expansion of $S(t;t_1)$, which is
\begin {equation}
   S(t;t_1) = (t-t_1) - \frac{1}{3!} \, \omega_0^2(t_1) \, (t-t_1)^3
            + O[(t-t_1)^5] ,
\end {equation}
so that
\begin {equation}
   C(t;t_1) = 1 - \frac{1}{2!} \, \omega_0^2(t_1) \, (t-t_1)^2
            + O[(t-t_1)^4] .
\end {equation}
I shall not need it, but the corresponding derivative of $C$ is
\begin {equation}
   \omega_0^2(t_1) \, S(t;t_1) = \partial_{t_1} C(t;t_1) .
\label {eq:Cderiv}
\end {equation}
Note that the relations (\ref{eq:Sderiv}) and (\ref{eq:Cderiv})
involve $t_1$ derivatives --- I will discuss the case of $t$
derivatives later.

Now substitute (\ref{eq:Sderiv}) into (\ref{eq:I2}):
\begin {equation}
  {\cal I}
  = - \lim_{t\to\infty}
    \ln \left[
       \frac{S(t;t_1)}
            {t-t_1}
    \right] \biggr|_{t_1=0}^{t_1=t} 
  = \lim_{t\to\infty}
    \ln \left[ \frac{S(t;0)}{t} \right]
  = \lim_{t\to\infty}
    \ln \left[ \partial_t S(t;0) \right] .
\end {equation}
Combining with (\ref{eq:result1}),
\begin {equation}
   \omega \, \frac{d}{d\omega}(I-I_{\rm vac}) =
   \frac{\alpha}{\pi} \, x \, P_{s{\to}{\rm g}}(x) \,
   \lim_{t\to\infty}
   \ln \bigl| \partial_t S(t;0) \bigr| .
\label {eq:result2}
\end {equation}


\subsection {Final simplification}

The result (\ref{eq:result2}) is perfectly adequate, but it is amusing
to put it in a final form that is even more closely analogous to the
result (\ref{eq:brick}) for the brick problem.

Note that any solution to a linear differential equation can
be written as a superposition of others.
So $S(t;t_1)$ and $C(t;t_1)$ can be expressed
as superpositions of $S(t;t_0)$ and $C(t;t_0)$ for any $t_0$.
Specifically,
\begin {align}
   S(t;t_1) &= C(t_1;t_0) \, S(t;t_0) - S(t_1;t_0) \, C(t;t_0) ,
\label {eq:Sbasis}
\\
   C(t;t_1) &= - \partial_{t_1} C(t_1;t_0) \, S(t;t_0)
               + \partial_{t_1} S(t_1;t_0) \, C(t;t_0) .
\label {eq:Cbasis}
\end {align}
To verify these formulas, one just needs to check the boundary
conditions.  The conditions
$S(t_1;t_1) = 0$ and $\partial_t C(t;t_1) |_{t=t_1} = 0$ are easy.
The other two, $\partial_t S(t;t_1) |_{t=t_1} = 1$ and
$C(t_1;t_1) = 1$, follow from the 
time independence of the Wronskian,
\begin {equation}
   \Bigl[
     C(t;t_0) \, \partial_t S(t;t_0) - S(t;t_0) \, \partial_t C(t;t_0)
   \Bigr]_{t=t_1}
   =
   \Bigl[
     C(t;t_0) \, \partial_t S(t;t_0) - S(t;t_0) \, \partial_t C(t;t_0)
   \Bigr]_{t=t_0}
   =
   1 .
\end {equation}

From (\ref{eq:Sbasis}), we see that $S(t;t_1)$ is
anti-symmetric in its arguments:
\begin {equation}
   S(t_2;t_1) = - S(t_1;t_2) .
\end {equation}
We can then combine this with (\ref{eq:Sderiv}) for the $t_1$ derivative
of $S$ to get a formula for the $t$ derivative:
\begin {equation}
   \partial_t S(t;t_1) = - \partial_t S(t_1;t) = C(t_1;t) .
\label {eq:Stderiv}
\end {equation}
[Eq. (\ref{eq:Cbasis}) does not allow us to deduce any comparable
symmetry property of $C$.]
We can now use (\ref{eq:Stderiv}) to rewrite (\ref{eq:result2})
in the form
\begin {equation}
   \omega \frac{d}{d\omega}(I-I_{\rm vac}) =
   \frac{\alpha}{\pi} \, x \, P_{s{\to}{\rm g}}(x) \,
   \ln\left| C(0;\infty) \right| ,
\end {equation}
which is Eq.\ (\ref{eq:wow}) of the introduction.


\section {Examples}
\label {sec:examples}

One can of course solve the differential equation (\ref{eq:c})
numerically for any desired time-dependence of $\qhat(t)$ along
the path of the particle.  In this section, I give a few examples
that have analytic solutions.

\subsection {The brick problem}

Consider the case where the particle travels distance $L$ through
a uniform medium and then emerges into vacuum.  So
\begin {equation}
   \omega_0^2(t) =
   \begin {cases}
      \omega_0^2, & t<L; \\
      0         , & t>L.
   \end {cases}
\end {equation}
The solution $c(t)$ to (\ref{eq:c}) is then
\begin {equation}
  c(t) =
  \begin {cases}
    \cos\bigl(\omega_0 (L-t)\bigr),
       & t < L; \\
    1,
       & t > L.
   \end {cases}
\end {equation}
Eq.\ (\ref{eq:wow}) then reproduces the result (\ref{eq:brick})
of BDMS \cite{BDMS}.

Using the fact that $\omega_0$ is proportional to $(-i)^{1/2}$,
one can alternatively write the result solely in terms of real
quantities using the identity
\begin {equation}
   \ln |\cos(e^{-i\pi/4}x)|
   = \half \ln\left[
       \half\cosh(\sqrt2 \, x) +
       \half\cos(\sqrt2 \, x) \right] .
   \qquad
   \mbox{($x$ real)}
\end {equation}
The large $L$ behavior is
\begin {equation}
   \ln |\cos(\omega_0 L)| \simeq \frac{|\omega_0| L}{\sqrt2} - \ln 2 ,
\label {eq:brickbulk}
\end {equation}
up to exponentially small corrections.  
(But you shouldn't take seriously the $\ln 2$ term because remember that
I've only treated $\omega_0$ itself up to leading-log order.)
In this limit, one can write
\begin {equation}
   \omega \frac{d}{d\omega}(I-I_{\rm vac}) \simeq
   \omega \frac{d\Gamma_{\rm bulk}}{d\omega} \, L ,
\end {equation}
with
\begin {equation}
   \omega \frac{d\Gamma_{\rm bulk}}{d\omega} \equiv
   \frac{\alpha}{\pi\sqrt{2}} \, x \, P_{s{\to}{\rm g}}(x) \,
   |\omega_0| .
\label {eq:Gammabulk}
\end {equation}

For fixed $x$, the small $L$ behavior is%
\footnote{
   Readers familiar with the fact that the medium-induced contribution
   to energy loss is proportional to $\qhat L^2$ for small $L$
   \cite{BDMPS2}
   may
   wonder how the $L^4$ behavior in the spectrum (\ref{eq:bricksmallL}) is
   consistent.  In (\ref{eq:bricksmallL}), the limit is that
   $L$ is small compared to the formation time, which
   is of order $(xE/\qhat)^{1/2} = (\omega/\qhat)^{1/2}$
   for $x$ not close to 1.  In contrast, the small $L$ formula
   for energy loss assumes $L \ll (E/\qhat)^{1/2}$.  In the latter limit,
   when the energy loss is determined by integrating
   $\omega \, dI/d\omega$ over $\omega$, the integral is dominated
   by $\omega$'s for which the formation time is of order $L$
   ($\omega \sim \qhat L^2$),
   where the small $L$ assumption of (\ref{eq:bricksmallL}) has
   just started to fail.
   Using (\ref{eq:bricksmallL}) merely as a parametric estimate
   then yields
   $\Delta E \sim \alpha \omega |\omega_0|^4 L^4
   \sim \alpha \omega (\qhat/\omega)^2 L^4
   \sim \alpha \qhat L^2$.
}
\begin {equation}
   \ln |\cos(\omega_0 L)| \simeq \tfrac1{12} (|\omega_0| L)^4 .
\label {eq:bricksmallL}
\end {equation}
Small $L$ in this context means $|\omega_0| L \ll 1$,
equivalent to $L \ll L_{\rm cr}$.
But keep in mind that the harmonic oscillator approximation
breaks down for calculations of the spectrum
when $L \lesssim L_{\rm cr}/[\ln(Q_\perp^2/\md^2)]^{1/2}$
\cite{N1}.


\subsection {Exponential Profile}
\label {sec:exp}

Consider an exponential profile
\begin {equation}
   \omega_0^2(t) =
   \omega_0^2(0) \, e^{-t/L} .
\end {equation}
The solution is
\begin {equation}
   c(t) = J_0\left(2\, \omega_0(0) \, L e^{-t/2L}\right) ,
\end {equation}
giving
\begin {equation}
   \omega \frac{d}{d\omega}(I-I_{\rm vac}) =
   \frac{\alpha}{\pi} \, x \, P_{s{\to}{\rm g}}(x) \,
   \ln\left| J_0\bigl( 2 \, \omega_0(0) L \bigr) \right| .
\label {eq:expresult}
\end {equation}


\subsection {Power Law Relaxation}

Motivated by modeling Bjorken expansion,
BDMS \cite{BDMSc} considered the case where $\qhat$ falls like a
power of time and then suddenly vanishes (the particle emerges into
vacuum) at time $L$.  So
\begin {equation}
   \omega_0^2(t) =
   \begin {cases}
      \omega_0^2(t_0) \, \left(\frac{t_0}{t}\right)^a ,
            & t_0 < t < t_0 + L ; \\
      0,    & t_0 + L < t ;
   \end {cases}
\end {equation}
where $a$ is some power and
I've now labeled the time of the initial hard process as
$t_0$ rather than zero.  The solution to (\ref{eq:c}) is
then $c(t) = 1$ for $t > t_0+L$ (the vacuum solution) and
\begin {align}
 c(t) &=
 \left(\frac{z}{z_L}\right)^\nu
 \frac{J_\nu(z) \, Y_{\nu-1}(z_L) - Y_{\nu}(z) \, J_{\nu-1}(z_L)}
      {J_\nu(z_L) \, Y_{\nu-1}(z_L) - Y_{\nu}(z_L) \, J_{\nu-1}(z_L)}
\nonumber\\
 &=
 \frac{\pi z_L}{2}
 \left(\frac{z}{z_L}\right)^\nu
 \bigl[ J_\nu(z) \, Y_{\nu-1}(z_L) - Y_{\nu}(z) \, J_{\nu-1}(z_L) \bigr]
 \qquad\qquad
 (t < t_0+L)
\label {eq:Cpower}
\end {align}
where%
\footnote{
   My $z$ differs by a factor of $i$ from that of Ref.\ \cite{BDMSc},
   which is why they have modified Bessel functions $K$ and $I$ instead
   of $J$ and $Y$.  Also, an equivalent way of writing (\ref{eq:Cpower})
   is to replace $Y_{\nu-1}$ and
   $Y_{\nu}$ by $J_{1-\nu}$ and $J_{-\nu}$.
   If comparing to Ref.\ \cite{BDMSc}, keep in mind that they solve
   a slightly different problem, as explained in footnote
   \ref{foot:problem}.
}
\begin {align}
   \nu &\equiv \frac{1}{2-a} \,,
\\
   z = z(t) &\equiv 2\nu \, \omega_0(t_0) \, t_0
       \left(\frac{t}{t_0}\right)^{1/2\nu} ,
\\
   z_0 &\equiv z(t_0) ,
\\
   z_L &\equiv z(t_0+L) .
\end {align}
The final result is then
\begin {equation}
   \omega \frac{d}{d\omega}(I-I_{\rm vac}) =
   \frac{\alpha}{\pi} \, x \, P_{s{\to}{\rm g}}(x) \,
   \ln\left|
     \left(\frac{t_0}{t_0+L}\right)^{1/2}
     \frac{J_\nu(z_0) \, Y_{\nu-1}(z_L) - Y_{\nu}(z_0) \, J_{\nu-1}(z_L)}
          {J_\nu(z_L) \, Y_{\nu-1}(z_L) - Y_{\nu}(z_L) \, J_{\nu-1}(z_L)}
   \right| .
\end {equation}


\subsection {\boldmath$\sech^2$ Profile}

As a final analytic example, consider a hard particle starting at
$t=t_0$ with profile
\begin {equation}
   \omega_0^2(t) =
   \Omega^2 \, \sech^2\left(\frac{t}{L}\right) .
\end {equation}
The solution is
\begin {equation}
   \omega \frac{d}{d\omega}(I-I_{\rm vac}) =
   \frac{\alpha}{\pi} \, x \, P_{s{\to}{\rm g}}(x) \,
   \ln\left| F\bigl(a_+,a_-;1;\frac{1}{e^{2t_0/L}+1}\bigr) \right| ,
\end {equation}
where $F$ is the hypergeometric function and
\begin {equation}
   a_\pm \equiv \half \pm \half\sqrt{1 + (2\Omega L)^2} .
\end {equation}


\section {General Solution: Limiting Cases}
\label {sec:limits}

I now turn to the behavior of the general solution (\ref{eq:wow})
for the limits of small or large width of the medium for fixed $x$.

\subsection{Small width}

In this case, we can solve the differential equation
\begin {equation}
   \ddot c(t) = - \omega_0^2(t) \, c(t)
\end {equation}
by perturbing around the vacuum solution $c_{\rm vac}(t) = 1$.
The solution is
\begin {equation}
   c(t) = 1 + c_1(t) + c_2(t) + O(\omega_0^3) ,
\end {equation}
where
\begin{align}
   c_1(t) &=
   - \int_t^\infty dt' (t'-t) \omega_0^2(t') ,
\\
   c_2(t) &=
   - \int_t^\infty dt' (t'-t) \omega_0^2(t') c_1(t') .
\end {align}
Now recall that $\omega_0^2$ is proportional to $-i$ so that
$c_1$ is imaginary and $c_2$ is real.  Then
\begin {align}
   \ln \left| c(0) \right|
   &\simeq
   \half \ln \left[ \bigl(1 + c_2(0)\bigr)^2
                    + \bigl|c_1(0)\bigr|^2 \right]
\nonumber\\
   &\simeq
   \half |c_1(0)|^2 + c_2(0)
\nonumber\\
   &\simeq
   \tfrac12 \left( \int_0^{\infty} dt' \ t' \, |\omega_0^2(t')| \right)^2
   - \int_0^\infty dt' \> t' |\omega_0^2(t)|
     \int_{t'}^\infty dt'' \> (t''-t') \, |\omega_0^2(t'')| .
\end {align}
This is the general form of the small-width answer, of which
(\ref{eq:bricksmallL}) is a specific case.


\subsection{Large width}

Now consider the case where $\omega_0^2(t)$ is a very slowly varying
function of $t$.  Then we can make an adiabatic approximation, and the
most important feature of the solution for $s(t)$ will be a ``phase
factor'' that is approximately%
\footnote{
  Because $\omega_0$ is proportional to $\exp(-i\pi/4)$,
  the other solution $\exp\left( -i \int_t^\infty dt' \omega_0(t') \right)$
  is, in the large width limit, exponentially small at $t=0$, and
  so its contribution to $c(0)$ can be neglected.
}
\begin {equation}
   c(t) \sim \exp\left( i \int_t^\infty dt' \> \omega_0(t') \right)
   = \exp\left( \frac{1}{\sqrt2} \int_t^\infty dt'\> |\omega_0(t')| \right)
     \exp\left( \frac{i}{\sqrt2} \int_t^\infty dt'\> |\omega_0(t')| \right)
\end {equation}
Neglecting prefactors (whose effect is parametrically
smaller than the exponent),
\begin {equation}
  \ln |c(0)| \simeq
  \frac{1}{\sqrt2} \int_0^\infty dt\> |\omega_0(t)|
  .
\end {equation}
Comparing to (\ref{eq:Gammabulk}), this gives
\begin {equation}
   \omega \frac{d}{d\omega}(I-I_{\rm vac}) =
   \int_0^\infty dt \> \omega \frac{d\Gamma_{\rm bulk}}{d\omega}(t) ,
\label {eq:adiabatic}
\end {equation}
as you would expect: In the limit of very thick, slowly varying media,
you just treat the problem as locally uniform, use the result for
the bremsstrahlung rate in an infinite, uniform medium, and
integrate.

There is a technical subtlety if one goes to next order in the
adiabatic expansion and looks at the prefactor.  The assumption of
the adiabatic expansion is that $|\dot\omega_0| \ll |\omega_0^2|$.
To first order in the prefactor, the solution for $C(t_1;t_2)$ is
\begin {equation}
   C(t;t_2) \simeq
   \left[ \frac{\omega_0(t_2)}{\omega_0(t)}  \right]^{1/2}
   \exp\left( i \int_t^{t_2} dt' \> \omega_0(t') \right) .
\end {equation}
Then
\begin {equation}
   \ln |c(0)|
   \simeq
   \frac{1}{\sqrt2} \int_0^\infty dt\> |\omega_0(t)|
   + \frac12 \ln \left| \frac{\omega_0(\infty)}{\omega_0(0)} \right| .
\label {eq:ad_problem}
\end {equation}
Since $\omega_0(\infty) = 0$, this answer suffers from a logarithmic
divergence.

The problem is that the adiabatic assumption
$|\dot\omega_0| \ll |\omega_0^2|$ must break down at sufficiently
late times.  As an example, consider the exponential distribution
of Sec.\ \ref{sec:exp}.  The adiabatic assumption first breaks down
when $|\omega_0|$ drops to $|\omega_0| \sim 1/L$.  If we use this
value of $\omega_0$ to cut off the logarithm in (\ref{eq:ad_problem}),
then we find a correction to the bulk result of size
$-\half \ln(|\omega_0(0)| L)$.
And in fact, exactly such a correction appears
in the large $L$ expansion of the exact result in (\ref{eq:expresult}),
which gives
\begin {equation}
   \ln\left| J_0\bigl( 2 \, \omega_0(0) L \bigr) \right|
   =
   \sqrt2 |\omega_0(0)| L - \half \ln\bigl(|\omega_0(0)| L\bigr) + O(1) .
\end {equation}


\section {Pair Production \boldmath${\rm g}\to{\rm q}\bar{\rm q}$}
\label {sec:pair}

Previous results are easily modified for the case of pair production
${\rm g}\to{\rm q}\bar{\rm q}$.  First, one uses the appropriate
DGLAP vacuum splitting function, so the overall result (\ref{eq:wow})
becomes
\begin {equation}
   x \frac{d}{dx}(I-I_{\rm vac}) =
   \frac{\alpha}{\pi} \, x \, P_{{\rm g}{\to}{\rm q}}(x)
   \ln\left| c(0) \right| ,
\end {equation}
with
\begin {equation}
   P_{{\rm g}\to {\rm q}}(x)
   = \Nf \tf [x^2+(1-x)^2]
\end {equation}
if one sums over all quark flavors.  Here $x$ is the momentum fraction
of the quark.  One must also change the factors in the definition
(\ref{eq:omega}) of $\omega_0^2$, as I shall discuss.
The only other change necessary is to appropriately change the
group factors in Eq.\ (\ref{eq:Gamma3}) for $\Gamma_3$
to reflect the different arrangement
of color representations in the splitting process from
${\rm F} \to {\rm AF}$ to ${\rm A} \to {\rm F\bar F}$.
The generalization of (\ref{eq:Gamma3}) is%
\footnote{
  For this form, see the discussion surrounding Eq.\ (6.11) and
  footnote 24 of Ref.\ \cite{AMYsansra}.
}
\begin {align}
  \Gamma_3(\B,t) = ~
    & \half (C_{R_2} + C_{R_3} - C_{R_1}) \, \bar\Gamma_2(x_1 \B,t)
\nonumber\\
  + & \half (C_{R_3} + C_{R_1} - C_{R_2}) \, \bar\Gamma_2(x_2 \B,t)
\nonumber\\
  + & \half (C_{R_1} + C_{R_2} - C_{R_3}) \, \bar\Gamma_2(x_3 \B,t)
\end {align}
for a $R_1 \to R_2 R_3$ splitting process with corresponding
momentum fractions
\begin {equation}
   x_1 = 1 , \qquad  x_2 = x, \qquad x_3 = 1-x .
\end {equation}
For $s \to {\rm g}s$ processes, this gives (\ref{eq:Gamma3}).
For ${\rm g}\to{\rm q}\bar{\rm q}$, the color factors of
(\ref{eq:Gamma3}) (or equivalently the momentum fractions)
are permuted to
\begin {equation}
  \Gamma_3(\B,t) =
  (\cf - \half\ca) \, \bar\Gamma_2(\B,t)
  + \half\ca \, \bar\Gamma_2(x\B,t)
  + \half\ca \, \bar\Gamma_2\bigl((1-x)\B,t\bigr) .
\end {equation}
The resulting value of $\omega_0^2$ replacing (\ref{eq:omega2}) is then
\begin {equation}
   \omega_0^2 =
   - i \,
   \frac{[\cf - x(1-x)\ca] \, \qhat}{2 x (1-x) E} \,.
\end {equation}


\section {Convergence of the opacity expansion}
\label{sec:opacity}

The opacity expansion investigated by Wiedemann \cite{Wopacity}
and Gyulassy, Levai, and
Vitev (GLV) \cite{GLV} involves analyzing bremsstrahlung in the
QCD medium by expanding order by order in the number of
elastic scatterings.
It is interesting to ask what happens if such an expansion is made
in a case where the leading-log calculation of BDMS is valid.
An expansion in powers of elastic collisions is equivalent to
an expansion in powers of $\Gamma_3$ (\ref{eq:Gamma3}), which
in leading-log approximation is equivalent to an expansion in
powers of $\omega_0^2$ (\ref{eq:omega2}).

Now consider BDMS's result (\ref{eq:brick}) for the brick problem,
and rewrite it in the form
\begin {equation}
   \omega \, \frac{d}{d\omega}(I-I_{\rm vac}) =
   \frac{\alpha}{2\pi} \, x \, P_{s{\to}{\rm g}}(x) \,
   \ln\Bigl[ \cos(e^{i\pi/4} z^{1/2}) \cos(e^{-i\pi/4}z^{1/2}) \Bigr],
\label {eq:brickz}
\end {equation}
where 
\begin {equation}
   z \equiv |\omega_0^2| L^2 .
\end {equation}
The opacity expansion of this result is its Taylor series in $z$,
proportional to
\begin {equation}
   \ln\Bigl[ \cos(e^{i\pi/4} z^{1/2}) \cos(e^{-i\pi/4}z^{1/2}) \Bigr]
   = \tfrac16 z^2
   - \tfrac{17}{1260} z^4
   + \tfrac{691}{467775} z^6
   - \cdots .
\label {eq:glvtest}
\end {equation}
Mathematically, the expression (\ref{eq:brickz}) is an analytic
function of $z$, and therefore its radius of convergence is given
by the distance to the nearest singularity in the complex $z$ plane.
The nearest singularities are the branch points of the logarithm
where either of the cosines vanish, at $z=\pm i (\pi/2)^2$.
In this example, the opacity expansion therefore only convergences
for $|z| < (\pi/2)^2$, which corresponds to
\begin {equation}
   L < \frac{\pi/2}{|\omega_0|} \,.
   \qquad \mbox{(brick)}
\end {equation}
Recall that, qualitatively, $1/|\omega_0|$ is of order the
formation time.  The conclusion is that the opacity expansion does
not converge when the medium is thicker than roughly the formation
time.

\begin{figure}[t]
\includegraphics[scale=1.50]{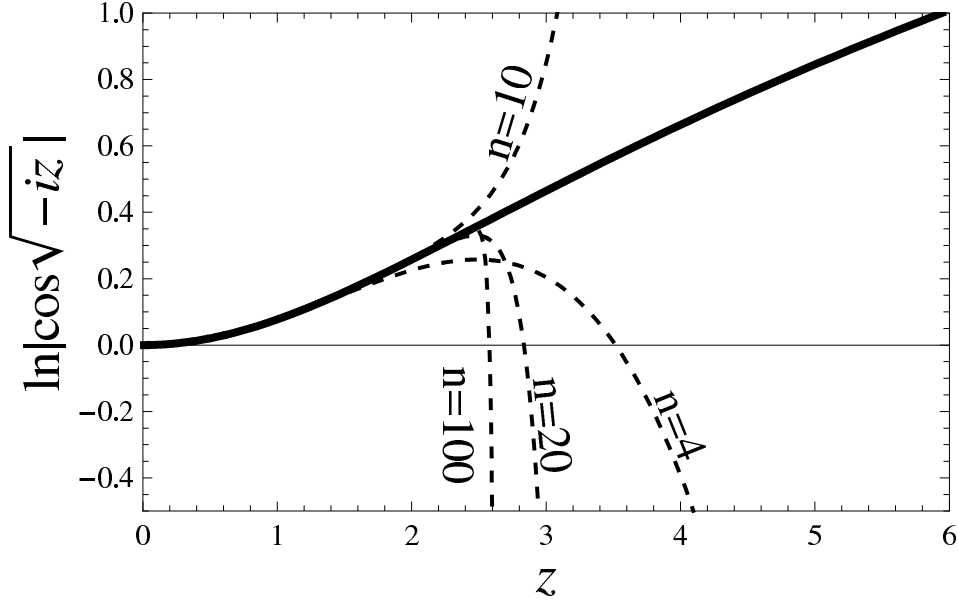}
\caption{%
    \label{fig:glv}
    The function of (\ref{eq:glvtest}) [solid line]
    vs $z = |\omega_0|^2 L^2$ compared to its Taylor series expansion
    to $n$th order for selected values of $n$ [dashed lines].
}
\end{figure}

In Fig.\ \ref{fig:glv}, I show the function (\ref{eq:glvtest}) vs its
expansion to $n$th order in the opacity expansion for several $n$.
One can see the failure of convergence beyond $z=(\pi/2)^2$.

One of the uses of the opacity expansion has been as a hook to
derive general results by summing up the expansion to all orders,
arriving at formalism related to BDMPS and Zakharov (for example,
as in Ref.\ \cite{Wopacity}).  In this case,
the lack of convergence of the Taylor series for large $L$
does not matter.

Readers may wonder at the juxtaposition of the opacity expansion and
the leading logarithm approximation.  In the large $N_{\rm coh}$ limit
of the leading logarithm approximation, $z=1$ in Fig.\ \ref{fig:glv}
represents a very large number of elastic scatterings.  But the
answer is nonetheless reproduced well by the $n{=}4$ curve, which only
includes up to four scatterings.  How can this be?  The reason is
that the LPM effect causes even a large number of scatterings to
behave like a single scattering if they occur within a distance
small compared to the formation time.  For this reason, it is possible
for just four scatterings, spread out across $L$, to reproduce the
same total bremsstrahlung rate as a large number of scatterings,
in leading log approximation.

I should clarify that the expansion discussed here depends on first
making the leading log approximation, treating $\qhat$ as a constant,
and only then making the opacity expansion.  So, for instance, I have
ignored the fact that the upper limit $Q_\perp$ of the logarithm in
(\ref{eq:qhatLL}) depends on the number $n$ of collisions.  In
particular, readers familiar with the opacity expansion may wonder at
the absence of a leading $n=1$ term in the expansion (\ref{eq:glvtest}),
proportional to $z$.  This is a special consequence of the leading-log
approximation \cite{ZakharovResolution,N1}.%
\footnote{
   Specifically, consider Eq.\ (6.7) of Ref.\ \cite{Wopacity}, using
   definitions (3.39), (5.6--9) and (5.11) of that reference.
   The leading-log approximation is
   $\bar\sigma(\rho) \propto \rho^2$, which corresponds to
   $\bar\Sigma(\q_\perp) \propto \nabla^2\delta^{(2)}(\q_\perp)$.
   If one uses this form of $\bar\Sigma$ and integrates
   Eq.\ (6.7) of Ref.\ \cite{Wopacity} over all
   bremsstrahlung gluon transverse momenta $\k_\perp$
   (making the $k_\perp \ll k$ approximation by integrating all the
   way up to $k_\perp = \infty$), one finds a zero result.
}

One might wonder whether the lack of convergence
is an artifact of the brick problem,
where $\omega_0^2(t)$ is not an analytic function of time.
However, one can draw the same conclusion from the exponential
profile (\ref{eq:expresult}).  In this case, the singularity
occurs at the first zero of the Bessel function, when its argument
is $2.40482\cdots$.
The corresponding condition for convergence of the opacity expansion
in this case is
\begin {equation}
   L < \frac{2.40482}{2 |\omega_0(0)|} \,.
   \qquad \mbox{(exponential)}
\end {equation}

These results have been derived in the leading-log approximation.
In situations where corrections to the leading-log approximation
are small, one expects similar conclusions since small perturbations
will not remove the presence of singularities.
The non-convergence of the opacity expansion might
possibly be related to the observed poor convergence 
in numerical results at small $x$ by Wicks, shown in Appendix B of
Ref.\ \cite{Wicks}, since $L/L_{\rm cr} \propto x^{-1/2}$ in
the small $x$ limit.




\begin{acknowledgments}

I am indebted to Guy Moore and Rudolph Baier for useful discussions.
I am also grateful to the referee for bringing my attention to
the failure of the harmonic oscillator approximation at sufficiently
small $L$.
This work was supported, in part, by the U.S. Department
of Energy under Grant No.~DE-FG02-97ER41027.

\end{acknowledgments}


\appendix

\section{Relation of Notation to Other Authors}
\label{app:notation}

\subsection{Zakharov}

The equations I give in Sec.\ \ref{sec:general} are organized slightly
differently than by Zakharov \cite{Zakharov2}.
I will compare my conventions specifically to Ref. \cite{Zakharov2}.
There, Zakharov ignores the fact that effective particle masses in a
non-uniform medium will depend on position and time.  If one treats
them as constant, then their contribution to the Hamiltonian
defined by (\ref{eq:H}), (\ref{eq:deltaE}) and (\ref{eq:Gamma3})
is an additive constant,
and their sole effect is to contribute a simple phase
$\exp[-i (\mbox{constant}) \, \Delta t]$ in the Green function, which
Zakharov explicitly factors out.  Specifically, the relationship between my
Hamiltonian and Green function and those of Zakharov (Z) 
\cite{Zakharov2} is
\begin {align}
  H^{\rm (my)} &= H^{\rm (Z)} + \frac{1}{L^{\rm (Z)}_f} \,,
\\
  G^{\rm (my)}(\B_2, t_2; \B_1, t_1) &=
  \exp\left[-\frac{i(t_2-t_1)}{L^{\rm (Z)}_f}\right]
  K^{\rm (Z)}(\B_2, t_2 | \B_1, t_1) ,
\end {align}
where
\begin {equation}
  L^{\rm (Z)}_f \equiv
 \frac{2 x (1-x) E}{x^2 \, m_s^2 + (1-x) \, m_{\rm g}^2}
\end {equation}
is what Zakharov calls the formation time.
Zakharov chooses to incorporate $1/L_f$ into his Hamiltonian
in the later work of Ref.\ \cite{BSZ}.

There is a difference between his and my use of the
phrase ``formation time.''
Zakharov uses it to mean the formation time in vacuum in the
case of exactly collinear bremsstrahlung, which is given by the
inverse of (\ref{eq:deltaE}) with $p_B$ set to zero.
I use it to mean the
formation time of typical bremsstrahlung in the medium, consistently
accounting for the LPM effect, which is the inverse of (\ref{eq:deltaE})
including the expectation of $p_B^2$.

The gradients in (\ref{eq:general}) correspond
(up to factors of $+i$ and $-i$)
to the operators $\p$ in Zakharov's definition of $g(\xi_1,\xi_2,x)$.

Finally, the way I have written Zakharov's three-parton and
dipole cross-sections $\sigma_3$ and $\sigma_2$
can be taken from BDMS's discussion of the equivalence of BDMPS
and Zakharov formalisms in Ref.\ \cite{BDMS}, as I shall discuss below.


\subsection {BDMS}

Throughout, where BDMS \cite{BDMS} or the earlier works of
BDMPS \cite{BDMPS1,BDMPS2,BDMPS3} expresses rates in terms of
density $\rho$ times a cross-section $\sigma$, I instead write
a rate $\Gamma$.  This allows one to more easily apply the formulas
to calculations that account for the dynamical nature of screening
in the plasma.

BDMS and BDMPS characterize the differential elastic cross-section
in terms of a normalized quantity
\begin {equation}
   V(Q^2) \equiv
   \frac{1}{\sigma_{\rm el}} \, \frac{d\sigma_{\rm el}}{d^2 Q} ,
\end {equation}
where they define $Q \equiv q_\perp/\md$.  Translating to the
language of rates, one may equivalently write
\begin {equation}
   V(Q^2) \equiv \frac{1}{\Gamma_{\rm el}}
   \, \frac{d\Gamma_{\rm el}}{d^2 Q} ,
\end {equation}
where $\Gamma_{\rm el}$ is written $\lambda^{-1} = \rho\sigma$
in the BDMPS formalism and $\lambda$ is the mean free path for
elastic collisions.
This expression is problematical for full, leading-order
perturbative calculations, however, because the total elastic
scattering rate $\Gamma_{\rm el}$ for a high-energy parton
traveling through a QCD plasma has a logarithmic infrared divergence in
perturbation theory, as can be seen by
integrating (\ref{eq:dsig}) over $d^2q_\perp$.
[The divergence does not appear in the discussions of BDMS and BDMPS
because, when they specialize to the case of Coulomb scattering, they
model $V(Q^2)$ as proportional to $1/(q_\perp^2+\md^2)^2$ rather than
the actual low-momentum perturbative behavior of (\ref{eq:dsig}).]
The divergence arises from the exchange of low-frequency
magnetic gluons, which are not screened, and is cut off only
by the non-perturbative physics of
magnetic confinement in hot QCD at a momentum scale $q_\perp \sim g^2 T$.
Formally, it is not clear whether there is any rigorous,
convention independent, non-perturbative definition of the total
rate $\lambda^{-1} = \Gamma_{\rm el}$, and so it is best to avoid
the quantity altogether.
Fortunately, this is just an issue of normalization convention.
The various quantities in the BDMPS formulas for the bremsstrahlung
rate appear in combinations where the factors of
$\lambda$ cancel, and I have chosen to avoid them in the formulas
of Sec.\ \ref{sec:general}.

A notational translation table is provided in Table \ref{tab:notation}.
The reasons for the complex conjugation that appears in some entries of
the BDMS column is that BDMS pick conventions where their analog of the
Schr\"odinger equation (\ref{eq:schro}) corresponds to a problem with
negative mass $M$.  One can convert to a Schr\"odinger equation with a
positive mass (Zakharov's convention, which I adopt) by taking the
complex conjugate of the equation, which takes $\psi \to \psi^*$,
$M \to -M$, and $\omega_0 \to \omega_0^*$.

\begin {table}
\begin {tabular}{|c|c|c|c|}
\hline
   this paper & BDMS \cite{BDMS} & Zakharov \cite{Zakharov2}
   & AMY \cite{AMYsansra} \\
\hline
  $\md$                    & $\mu$
     & & $\md$                                                         \\
  $\q_\perp$               & $\mu \underline{Q}$
     & & $\q_\perp$                                                    \\[4pt]
  $\bar\Gamma_{\rm el}$    & $\frac{1}{\lambda\cf} = \frac{\rho\sigma}{\cf}$
                                                                    && \\[4pt]
  $\frac{d\bar\Gamma_{\rm el}}{d^2q_\perp}$
                           & $\frac{V(Q^2)}{\mu^2 \lambda\cf}$
     &
     & $\frac{g^2}{(2\pi)^2}
         \int\!\frac{dq^z}{2\pi} \bigl\langle\!\!\bigl\langle
         A^-(Q) [A^-(Q)]^*
         \bigr\rangle\!\!\bigr\rangle_{\kern-3pt q^0=q^z}$             \\
  $\p_B$                   & $\mu(\underline{U}-x\underline{V})$
     & $\p$
     & $\h/p'$                                                         \\
  $\B$                     & $\underline{B}/\mu$
     & ${\bm\rho}$                                                   & \\[4pt]
  $\qhat$
                           & $\frac{\mu^2 \tilde v(0)}{\cf\lambda}$
     & $\tfrac32 n \, C_2(0)$                                        & \\[4pt]
  $\omega_0$               & $\frac{\omega_0^* \Nc}{2\cf\lambda}$   && \\[4pt]
  $t$                      & $\frac{2\cf\lambda\tau}{\Nc}$                  
     & $z$ or $\xi$                                                  & \\[4pt]
  $M$
     & $\frac{\tilde\kappa\Nc}{2\lambda\cf}
        = -\frac{2\lambda\cf \mu^2}{\Nc} m$
     & $\mu(x)$                                                      & \\[4pt]
  $G(\B_2,t_2;\B_1,t_1)$
     & $\mu^2 G^*(\underline{B}_2,z_2;\underline{B}_1,z_1)$
     & $\substack{e^{-i(\xi_2-\xi_1)/L_f}\\
        \qquad \times K({\bm\rho}_2, \xi_2 | {\bm\rho}_1, \xi_1)}$   & \\[4pt]
  $\bar\Gamma_2(\B)$
     & $\frac{1-\tilde V(\underline{B})}{\lambda\cf}$
     & $\tfrac38 \, n \, \sigma_2({\bm\rho})$                        & \\
\hline
\end {tabular}
\caption
    {
    Translation between notation of this paper and various authors.
    The entries in the $\qhat$ line are logarithmically divergent in
    the ultraviolet and should be understood as appropriately
    cut off for a leading-log approximation, as discussed in the text,
    or equivalently evaluated at some small effective value of $B$
    (BDMS) or $\rho$ (Zakharov) of order $1/Q_\perp$.
    \label {tab:notation}
    }
\end {table}

In BDMS \cite{BDMS}, the quark and gluon masses are ignored.  This is
parametrically valid when $Q_\perp \gg m_{\rm q}$ and $m_{\rm g}$,
which for a {\it thick} medium ($L \gtrsim L_{\rm cr}$)
corresponds to the high-energy limit $E \gg m_R^4/\qhat$.  In
perturbation theory, where $m_{\rm q} \sim m_{\rm g} \sim gT$,
this condition is
parametrically $E \gg T$.
However, in applications of the LPM effect where $E \sim T$
is of interest (such as leading-order calculations of viscosity
and other transport coefficients \cite{AMYx}), one should include the
mass terms.

Finally, there is an overall minus sign difference between my
(\ref{eq:general}) and the comparable
Eq.\ (59) of Ref.\ \cite{BDMS}.  One quick
way to resolve minus sign issues is to check that the final answer
for the effect
of the medium is positive in the limit of a very thick medium,
as in (\ref{eq:adiabatic}).%
\footnote{
  There appears to be a lost minus sign in the transition from
  Eqs. (31) and (33) to (51) of Ref.\ \cite{BDMS}, which then
  propagates to their (59).
}


\subsection {AMY}

Next, I wish to make contact with the notation used in
my previous work with Moore and Yaffe \cite{AMYsansra,AMYkinetic,AMYx}.
That analysis was for the case of an infinite, uniform, time-independent
medium.  Following Migdal \cite{Migdal}, one can treat this case by starting
with the non-vacuum part of
(\ref{eq:general}), changing integration variables from
$t_2$ to the time difference $\Delta t \equiv t_2-t_1$,
and then using time invariance to note that the Green function
depends only on $\Delta t$.
The $t_1$ integral then just gives a factor of the total time,
and the resulting
equation for the bremsstrahlung rate is
\begin {equation}
   \omega \, \frac{d\Gamma_{\rm brem}}{d\omega} =
   \frac{\alpha x \, P_{s{\to}{\rm g}}(x)}{[x(1-x) E]^2} \,
   \Real
   \int_0^\infty d(\Delta t) \>
   \Bigl[
     \grad_{\B_1} \cdot \grad_{\B_2}
       G(\B_2,\Delta t;\B_1,0)
   \Bigr]_{B_1=B_2=0} .
\label {eq:rate}
\end {equation}
Now define
\begin {equation}
   \f(\B,t) = 2i \Bigl[ \grad_{\B_1} G(\B,t;\B_1,0) \Bigr]_{B_1=0} ,
\end {equation}
where the overall normalization of $2i$ is chosen to make
contact with AMY conventions.
Each component of $f$ satisfies the same Schr\"odinger equation
(\ref{eq:schro}) that
$G$ does, so that
\begin {equation}
   i \partial_t \f(\B,t) = H \, \f(\B,t)
\label {eq:fschro}
\end {equation}
with initial condition
\begin {equation}
   \f(\B,0) = -2i \grad_{\B} \delta^{(2)}(\B) .
\end {equation}
Now define the time-integrated amplitude
\begin {equation}
   \f(\B) \equiv \int_0^\infty dt \> \f(\B,t) .
\end {equation}
Integrating both sides of (\ref{eq:fschro}) over time
(and noting that $\f(\B,t)$ decays with time because of the
$-i \Gamma_3$ piece of $H$),
\begin {equation}
   - 2 \grad_{\B} \delta^{(2)}(\B)
   = H \, \f(\B) .
\end {equation}
The rate (\ref{eq:rate}) can be written in terms of $\f(\B)$ as
\begin {equation}
   \omega \, \frac{d\Gamma_{\rm brem}}{d\omega} =
   \frac{\alpha x \, P_{s{\to}{\rm g}}(x)}{[x(1-x) E]^2} \,
   \Real
   \Bigl[
     (2i)^{-1} \grad_{\B} \cdot \f(\B)
   \Bigr]_{B=0} .
\label {eq:almostAMYrate}
\end {equation}
Now Fourier transform from $\B$ to $\p_B$.
Using the form (\ref{eq:H}) of $H$, the equation for
$\f$ becomes
\begin {multline}
   -2i\p_B
   = \delta E(p_B) \, \f(\p_B)
   - i \int d^2q_\perp \>
     \frac{d\bar\Gamma_{\rm el}}{d^2q_\perp}
     \biggl\{
       \half\ca \Bigl[ \f(\p_B) - \f(\p_B+\q_\perp) \Bigr]
\\
       + (C_s - \half \ca) \Bigl[ \f(\p_B) - \f(\p_B+x\q_\perp) \Bigr]
       + \half\ca \Bigl[ \f(\p_B) - \f\bigl(\p_B+(1-x)\q_\perp\bigr) \Bigr]
     \biggr\} .
\label{eq:almostAMY}
\end {multline}
Instead of $\p_B$, AMY uses the variable $\h \equiv \p_B P$.
In the case of bremsstrahlung, they define the momenta of the
splitting particles as $p'=P$, $k=xP$, and $p=(1-x)P$.
If one defines
\begin {equation}
   \F(\h) = P \f(\h/P) ,
\end {equation}
then (\ref{eq:almostAMY}) becomes
\begin {multline}
   -2i\h
   = \delta E \, \F(\h)
   - i \int d^2q_\perp \>
     \frac{d\bar\Gamma_{\rm el}}{d^2q_\perp}
     \biggl\{
       \half\ca \Bigl[ \F(\h) - \F(\h+p'\q_\perp) \Bigr]
\\
       + (C_s - \half \ca) \Bigl[ \F(\h) - \F(\h+k\q_\perp) \Bigr]
       + \half\ca \Bigl[ \F(\h) - \F(\h+p\q_\perp) \Bigr]
     \biggr\} .
\end {multline}
This is equation (6.7) of Ref.\ \cite{AMYsansra} if one
changes integration variable from $\q_\perp$ to $-\q_\perp$ in some
of the terms and recognizes that
\begin {equation}
     \frac{d\bar\Gamma_{\rm el}}{d^2q_\perp} =
     \frac{g^2}{(2\pi)^2} \, {\cal A}(q_\perp)
     \equiv
     \frac{g^2}{(2\pi)^2}
         \int_{-\infty}^{+\infty}\frac{dq^z}{2\pi}
         \bigl\langle\!\!\bigl\langle
         A^-(Q) [A^-(Q)]^*
         \bigr\rangle\!\!\bigr\rangle_{q^0=q^z} .
\end {equation}
With the same notation, the rate (\ref{eq:almostAMYrate}) becomes
\begin {align}
   \omega \, \frac{d\Gamma_{\rm brem}}{d\omega} =
   x \, \frac{d\Gamma_{\rm brem}}{dx} &=
   \frac{\alpha x \, P_{s{\to}{\rm g}}(x)}{4[x(1-x) E]^2} \,
   \int \frac{d^2 p_B}{(2\pi)^2} \>
   \Real
   [2\p_B \cdot \f(\p_B)]
\nonumber\\
   &=
   \frac{\alpha x \, P_{s{\to}{\rm g}}(x)}{4x^2(1-x)^2 E^6} \,
   \int \frac{d^2 h}{(2\pi)^2} \>
   \Real
   [2\h \cdot \F(\h)] .
\end {align}
This formula can be extracted from the rates per unit volume presented
for kinetic theory in AMY Ref.\ \cite{AMYkinetic}, for example,
with
\begin {equation}
   \gamma_{s\to gs}(E;xE,(1-x)E) =
   \frac{d_s \alpha \, P_{s{\to}{\rm g}}(x)}{(2\pi)^3 2x^2(1-x)^2 E^5} \,
   \int \frac{d^2 h}{(2\pi)^2} \>
   \Real
   [2\h \cdot \F(\h)] .
\end {equation}
More simply, if final-state factors of
$[1\pm f(xE)][1\pm f((1-x)E]$ are included, it corresponds to
Eq.\ (5) of Jeon and Moore \cite{JeonMoore} or Eqs.\ (1.1) and (4.1--2)
of Ref.\ \cite{ArnoldDogan}.
Jeon and Moore use the symbol $d\Gamma/dt$ to denote rate rather than
$\Gamma$.

Readers comparing to AMY should beware that AMY uses
the symbol $\Gamma$ to indicate
the rate per unit volume, integrating what I call $\Gamma_{\rm brem}$
over the initial particle's momentum with a factor of its
distribution function $f$ and including final state factors.%
\footnote{
   A pernicious factor of 2 that arises when comparing to AMY
   expressions is that they sum formulas for
   splitting of particle types $a \to bc$ over the types
   $b$ and $c$.  For bremsstrahlung from a quark, this gives rise
   to a factor of 2 because both ${\rm q}\to {\rm qg}$ and
   the identical ${\rm q}\to {\rm gq}$ are summed over.
   For ${\rm g}\to{\rm gg}$
   there is no such factor of 2, accounting for the relative factor
   of $1/2$ one needs to include when integrating over the final
   momentum fractions of two identical particles.
}


\subsection {Wiedemann}

Finally, I will translate to the notation of Wiedemann and
collaborators \cite{Wopacity,Wiedemann2,SalgadoWiedemann,KovnerWiedemann}
as presented in Salgado and
Wiedemann \cite{SalgadoWiedemann}.
They specialize to the $x\ll 1$ limit of soft bremsstrahlung gluons, but they
study more properties of the process, such as the angle between
the emitted gluons and the high-energy parton, and what happens when
the gluon momentum is so small that the approximation $k_\perp \ll k$
is no longer valid.  The basic result, Eq.\ (2.1) of
Ref.\ \cite{SalgadoWiedemann}, is
\begin {align}
   \omega \, \frac{dI}{d\omega}
   = &
   \frac{\alpha_s C_R}{(2\pi)^2\omega^2} \,
   2 \Real \int_{\xi_0}^\infty dy_l \int_{y_l}^\infty d\bar y_l
   \int d{\bm u} \int_0^{\chi\omega} d\k_\perp e^{-i\k_\perp\cdot{\bm u}}
\nonumber\\
   &\times
   e^{-(1/2)\int_{\bar y_l}^\infty d\xi \, n(\xi) \, \sigma({\bm u})}
   \frac{\partial}{\partial{\bm y}} \cdot \frac{\partial}{\partial{\bm u}}
\nonumber\\
   &\times
   \int_{{\bm y}=0=\r(y_l)}^{{\bm u}=\r(\bar y_L)} {\cal D}\r
   \exp\left[
     i\int_{y_l}^{\bar y_l}d\xi \> \frac{\omega}{2}
     \left( \dot\r^2-\frac{n(\xi)\,\sigma(\r)}{i\omega} \right)
   \right] .
\end {align}
The limit $k_\perp \le \chi \omega$ is used to restrict attention to
gluon bremsstrahlung in a finite opening angle $\Theta$ with
$\chi = \sin\Theta$.  In this paper, I have put no such restriction,
and I have assumed $k$ sufficiently large that
$k_\perp \ll k$ dominates.  This corresponds to
replacing the upper limit $\chi\omega$ on the $k_\perp$ integration
by infinity.  That integral then generates a factor of
$\delta^{(2)}({\bm u})$, which makes the ${\bm u}$ integration trivial.
Using the fact that their definition of $\sigma({\bm u})$ has
$\sigma(0)=0$, one then obtains
\begin {align}
   \omega \, \frac{dI}{d\omega}
   = &
   \frac{\alpha_s C_R}{\omega^2} \,
   2 \Real \int_{\xi_0}^\infty dy_l \int_{y_l}^\infty d\bar y_l
\nonumber\\
   &\times
   \frac{\partial}{\partial{\bm y}} \cdot \frac{\partial}{\partial{\bm u}}
   \int_{{\bm y}=\r(y_l)}^{{\bm u}=\r(\bar y_L)} {\cal D}\r
   \exp\left[i\int_{y_l}^{\bar y_l}d\xi \> \frac{\omega}{2}
   \left( \dot\r^2-\frac{n(\xi)\,\sigma(\r)}{i\omega} \right)
   \right] \biggl|_{{\bm u}={\bm y}=0}.
\end {align}
This is the small $x$ approximation to (\ref{eq:general}) of this
paper, with the notational translations shown in table
\ref{tab:Wiedemann}, my convention $\xi_0=0$,
the bremsstrahlung gluon mass ignored,
and the Green function expressed as a path integral.

\begin {table}
\begin {tabular}{|c|c|}
\hline
   this paper & Salgado \& Wiedemann \cite{SalgadoWiedemann}      \\
\hline
  $t_1$                    & $y_l$                                \\
  $t_2$                    & $\bar y_l$                           \\
  $\B_1$                   & ${\bm y}$                            \\
  $\B_2$                   & ${\bm u}$                            \\
  $\B$                     & $\r$                                 \\
  $\qhat$                  & $\hat q/\ca$                         \\
  $M$                      & $\omega$                             \\
  $\omega_0$
      & $\Bigl[(1+i)\sqrt{\frac{\hat q}{4\omega}}\Bigr]^*$        \\
  $\omega_0 L$
      & $\Bigl[(1+i)\sqrt{\frac{\omega_c}{2\omega}}\Bigr]^*$      \\
  $\Gamma_3(\B)$           & $\frac{n \, \sigma(\r)}{2}$          \\
  $\bar\Gamma_2(\B)$       & $\frac{n \, \sigma(\r)}{2\ca}$       \\
\hline
\end {tabular}
\caption
    {
    Translation between notation of this paper and Salgado and
    Wiedemann \cite{SalgadoWiedemann}, which studies the $x \ll 1$
    limit.
    \label {tab:Wiedemann}
    }
\end {table}


\end{document}